\documentclass[fleqn,usenatbib]{mnras}

\usepackage{newtxtext,newtxmath}

\usepackage[T1]{fontenc}

\DeclareRobustCommand{\VAN}[3]{#2}
\let\VANthebibliography\thebibliography
\def\thebibliography{\DeclareRobustCommand{\VAN}[3]{##3}\VANthebibliography}

\usepackage{graphicx}	
\usepackage{amsmath}	

\newcommand{\gaia}{\textit{Gaia }}

\title[M–-R Relation in OG-NG Systems]{Do Outer Giants Inflate Neptune-sized Planets? \\An Architecture-Dependent Mass--Radius Relation}

\author[D. Bashi]{
Dolev Bashi$^{1}$\thanks{E-mail: db975@cam.ac.uk}\\
$^{1}$Astrophysics Group, Cavendish Laboratory, University of Cambridge, JJ Thomson Avenue, Cambridge CB3 0HE, UK\\
}

\date{Accepted 2025 September 12. Received 2025 August 20; in original form 2025 July 9}

\pubyear{\the\year{}}

\begin{document}
\label{firstpage}
\pagerange{\pageref{firstpage}--\pageref{lastpage}}
\maketitle

\begin{abstract}
Exoplanet demographics increasingly reveal that planetary properties depend not only on local irradiation and composition but also on the wider system architecture. We analyse a sample of Neptune-sized short-period planets with well-measured masses and radii, identifying those whose host stars harbour at least one confirmed outer-giant (OG) companion. On the mass–radius (M--R) plane, the two populations diverge modestly: inner planets in OG systems cluster at systematically larger radii than their counterparts in no-giant (NG) systems, a result that remains suggestive after controlling for planet and stellar properties. 
Bayesian modelling quantifies the offset, revealing an average radius enhancement of $17\pm 4 \,\%$ for inner planets in OG systems relative to NG systems at fixed mass. Alternative cuts, including the use of a homogeneous set of parameters, confirm the robustness of the signal, though the result still relies on small-number statistics. Possible mechanisms for the observed inflation include prolonged inner-disc gas supply that boosted envelope accretion, and volatile enrichment by the outer giant.
If upheld, this empirical link between outer giants and inflated inner-planet radii offers a new constraint on coupled formation and evolution in planetary systems.
\end{abstract}




\section{Introduction}

The mass--radius (M--R) relation is the fundamental diagnostic linking an exoplanet’s bulk composition, internal structure, and thermal history to its observed mass and radius \citep{WeissMarcy14, Bashi17, ChenKipping17, Ulmer-Moll19, Otegi20, Edmondson23, Muller24}. 
Viewed demographically, the M--R plane can reveal whether system architecture leaves measurable fingerprints on planet properties.   

Beyond individual planets, multi-planet systems display striking internal regularities: neighbours tend to have nearly equal radii and dynamically regular spacings, the so-called `peas in a pod' pattern \citep{Weiss18, MillhollandWinn21, LozovskyPerets25}. Quantitative metrics have been developed to compare complete system architectures and confirm that most \emph{Kepler} multiplanets occupy a narrow locus in the period--radius space \citep{GilbertFabrycky20, BashiZucker21, HeWeiss23, Rice24}. These findings imply a strong memory of disc conditions, but also raise the question of how outer companions might disturb or modify such architectures \citep[e.g.,][]{Sobski23}.

Large-scale radial-velocity follow-up has begun to map the conditional occurrence of outer gas giants given the presence of close-in small planets. While some surveys report a positive correlation \citep{Rosenthal22, Bryan24, VanZandt25}, others find no statistically significant excess \citep{Bonomo23}. 

Population-synthesis and $N$-body models predict several pathways by which an outer giant can influence inner small planets \citep{BitschIzidoro23, Best24, Weber24}. Large-scale synthesis efforts classify such systems as a distinct mixed-architecture in which low-mass and giant planets form simultaneously \citep{Emsenhuber23}. Dynamical studies also link the eccentricity distribution of cold giants with volatile delivery and potential atmospheric growth for interior planets \citep{Kane24}.

Despite these advances, no study has directly compared the M--R relations of inner planets in systems with and without known outer giants. Here we fill that gap: using a uniform sample of transiting planets with precise masses and radii, we argue that inner Neptune-sized planets accompanied by an outer giant are larger by $\sim17$\,\% at fixed mass. This architecture-dependent offset adds a new dimension to the exoplanet M--R relation and provides a stringent constraint for formation models.

\section{Sample Selection}
\label{sec:Sample}

Our analysis is based on the NASA Exoplanet Archive `Planetary System' Table\footnote{\url{https://exoplanetarchive.ipac.caltech.edu}} \citep{NASA_Archive} which includes self-consistent planet and stellar parameters for a given planet. The snapshot used here was downloaded on 17 May 2025 and contained $5,903$ confirmed exoplanets.

\begin{figure}
   \centering
   \includegraphics[width=0.45\textwidth] {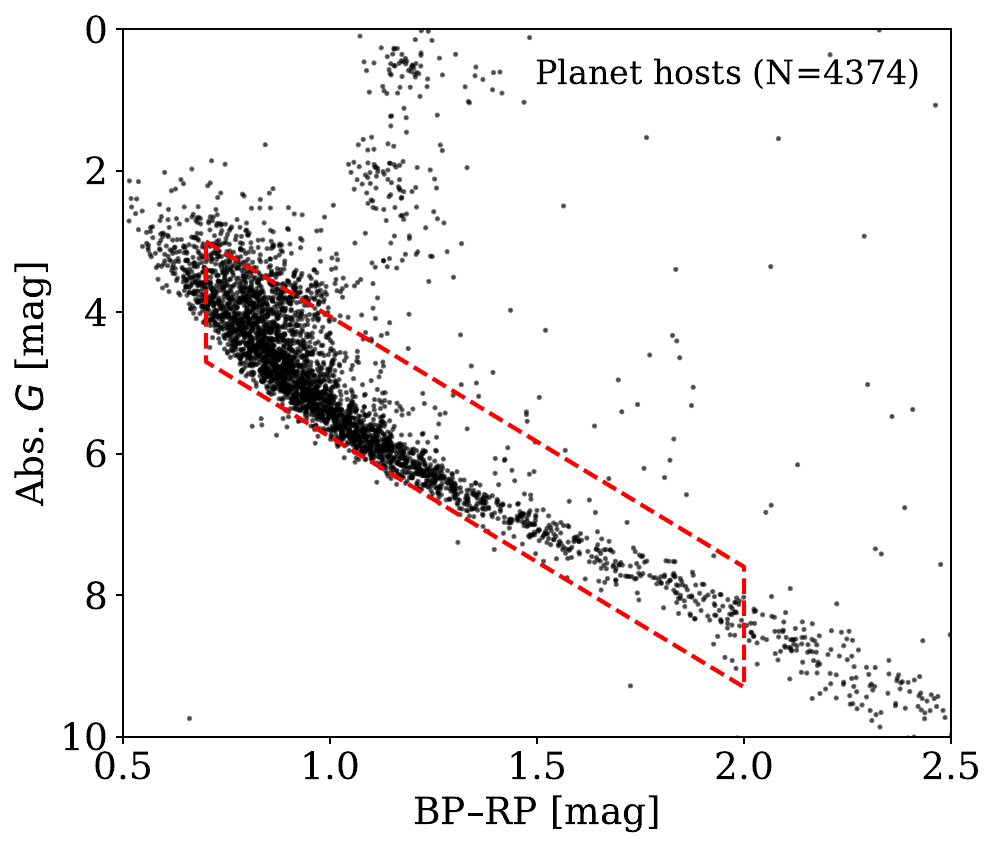}
   \caption{\textit{Gaia} CMD for the 4,374 confirmed planet hosts used to define the stellar sample. The red dashed polygon encloses the unevolved FGK main-sequence locus, removing evolved or late-type stars. The 3,952 planets (2,838 hosts) that survive this cut form the parent population for all subsequent analysis.} 
              \label{fig:CMD}%
    \end{figure}

To minimise systematics arising from stellar evolution, we apply a similar selection as used in \cite{Bashi24}, restricting the sample to unevolved FGK dwarfs using \textit{Gaia} DR3 photometry \citep{GaiaDR3} on a colour–magnitude diagram (CMD). Absolute $G$ magnitudes (\texttt{Abs. G}) were computed from the catalogue parallax, and we adopted \texttt{BP- RP} colours available for $4,374$ planet host stars ($5,860$ planets). Planets whose hosts lie inside the empirical polygon shown in Fig.\ref{fig:CMD} (red dashed line) defined by the vertices 
$(0.70, 3)$, $(2.0, 7.6)$, $(2.0, 9.3)$, $(0.70, 4.7)$, $(0.70, 3)$
in the (\texttt{BP- RP, Abs. G}) plane, were retained. The cut removes late-type stars, sub‑giants and giants while keeping main‑sequence stars roughly between spectral types F and K. While the exact polygon vertices selection might be somewhat arbitrary, we made sure that this selection does not change our final results. After this step, $3,952$ planets orbiting $2,838$ stars remained.

Our `inner-planet' sample is defined by a dynamical criterion, i.e., orbital period $P< 50$ days, which simply tags these objects as the close-in companions whose properties we wish to compare. To ensure that both mass and radius are derived homogeneously, we keep only planets with confirmed transits (\texttt{tran\_flag}=1) and radial-velocity (RV) detections (\texttt{rv\_flag}=1). Systems whose masses rely solely on transit-timing variations (TTV) are excluded because these planets, being preferentially at longer periods, appear systematically less massive than RV-detected planets of comparable size \citep{MillsMazeh17, Otegi20}. 

To minimise biases in stellar and planetary parameters, we exclude known multi-stellar systems (\texttt{sy\_snum}>1) and require good single-star astrometry from \textit{Gaia} DR3 (\texttt{RUWE}\,$\leq$\,1.4; \citealt{Lindegren21}). These cuts reduce the likelihood that unresolved companions dilute transit depths and bias planet radii (e.g. \citealt{Han25}). For RV-derived masses, we adopt values from the literature. While differences in eccentricity treatment and activity modelling can affect the inferred masses (e.g. \citealt{Osborne25}), we note this residual heterogeneity and interpret our population-level results with these caveats in mind.

We also exclude host stars cooler than $T_{\mathrm{eff}} = 4600\,\mathrm{K}$, as late-K and M dwarfs rarely host outer Jovian companions \citep{Sabotta21, Bryant23} and, owing to their faintness, introduce heterogeneous detection biases.

We restrict the radius to select mainly Neptune-sized planets with $R_{\mathrm p} \leq 6R_{\oplus}$. The lower bound is not a fixed cutoff, but instead varies with orbital period and stellar mass, following the location of the radius valley \citep{Fulton17, VanEylen18, HoVanEylen23, Rogers25}. Specifically, we require planets to lie above the empirical valley relation of \citet{HoVanEylen23} (see their Equation 11), with a small buffer of $5 \%$ to avoid including objects within the valley itself. In any case, to further remove highly irradiated planets whose atmospheres can be bloated by stellar heating, we also require a planetary equilibrium temperature, $T_{\mathrm{eq}}$, cooler than $1200\,\mathrm{K}$. These radius boundaries remove planets whose envelopes are likely stripped, while the upper bound excludes gas-giant analogues whose structure follows a different $M$–$R$ relation. 

Finally, we retain only those planets whose radii and masses are measured to better than 10\% and 20\%, respectively, so that model fits are not dominated by measurement noise. These cuts yield a well-characterised set of $42$ close-in Neptune-sized planets around $35$ stars suitable for testing whether outer giants inflate their envelopes.

Next, we looked for known giant planets among these systems. We define an outer giant as any confirmed companion with $M_{\mathrm p} > 80 M_{\mathrm \oplus}$ and an orbital period $P=300-10,000$ days. The same catalogue lists $241$ such planets. Host names were cross‑matched to label each inner‑planet system as either (i) an `OG' (outer‑giant) system when at least one outer giant is present, or (ii) an `NG' (no‑giant) system otherwise. This procedure yielded $8$ inner planets in $7$ OG systems and $34$ planets in $28$ NG systems. Importantly, our definition of `NG' does not imply the absence of an outer giant planet; rather, such companions may exist but remain undetected due to a limited RV search cadence.


A complete list of all inner planets in the OG and NG
subsamples, together with their radii and masses, are provided in
Appendix~\ref{app:inner_catalogue} (Tables~\ref{tab:OG}, \ref{tab:NG}).

\section{Inflation of Neptune-sized Planets: A Signature of Outer Giants?}
\label{sec:Results}

\begin{figure*}
   \centering
\includegraphics[width=12cm] {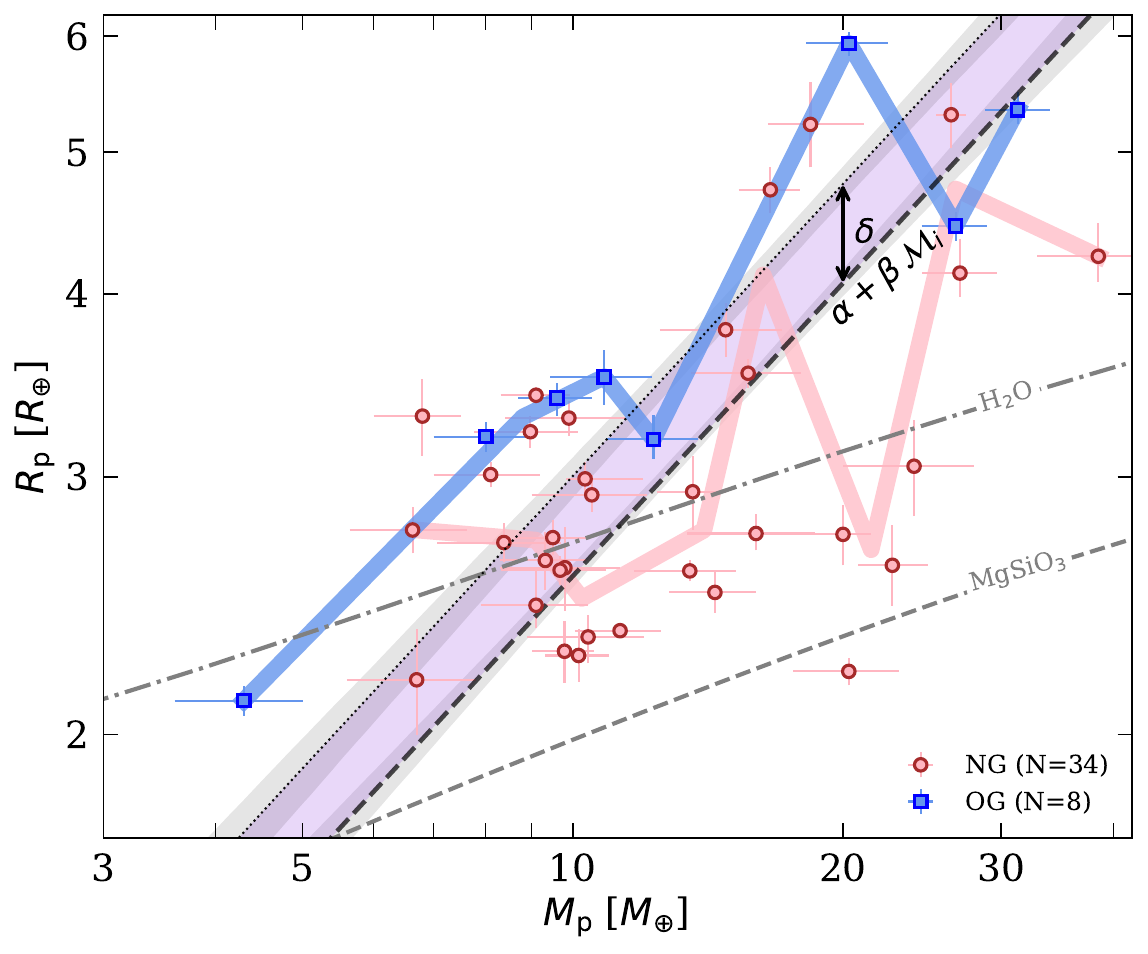}
   \caption{M--R diagram for inner ($P< 50$ days) Neptune-sized planets. Red circles represent inner planets in systems with no detected giant (NG), and blue squares are inner planets in systems with at least one outer ($P=300-10,000$ days) giant ($M > 80 M_{\oplus}$; OG). Blue and red bands indicate the median radii in mass bins for each subsample. The black dashed line shows the best-fit (posterior) linear trend $\alpha + \beta \mathcal{M}_i$, with the shaded purple region marking the offset $\delta$ associated with OG systems. Shaded grey bands illustrate the typical uncertainties of the fit. 
For reference, the faint grey lines indicate theoretical composition models of pure H$_{\mathrm{2}}$O ice and pure MgSiO$_{\mathrm{3}}$ rock.} 
              \label{fig:MR}%
    \end{figure*}

We first examine the overall M--R distribution of the inner planets in our sample. Fig.~\ref{fig:MR} shows a scatter plot of the planetary radius ($R_{\rm p}$) vs. mass ($M_{\rm p})$ for the two populations: red circles denote planets in NG systems, and blue squares denote planets in OG systems. Over-plotted on Fig.~\ref{fig:MR} are the theoretical composition model lines \citep{Zeng2019} of pure water ice (H$_{\mathrm{2}}$O) and pure rock (MgSiO$_{\mathrm{3}}$). The two samples exhibit the familiar broad scatter in radii for a given mass, reflecting diverse compositions from sub-Neptunes to Super-Neptune-sized planets. A visual trend is nonetheless apparent: for a given mass, blue squares (OG systems) tend to lie slightly above the red circles (NG systems). To guide the eye, we plot median radii in mass bins as blue and red bands. Based on this qualitative comparison, inner planets that coexist with outer giants appear to have somewhat larger radii, on average, than their counterparts without a known outer giant planet. 

Using a Mann-Whitney U-test, which directly compares whether planets in OG systems lie systematically farther above the 100\% H$_2$O M--R line than those in NG systems, we find a p-value of $0.00548$, providing significant evidence that OG planets are elevated relative to this theoretical H$_2$O line compared with their NG counterparts.

\begin{figure*}
   \centering
   \includegraphics[width=0.8\textwidth] {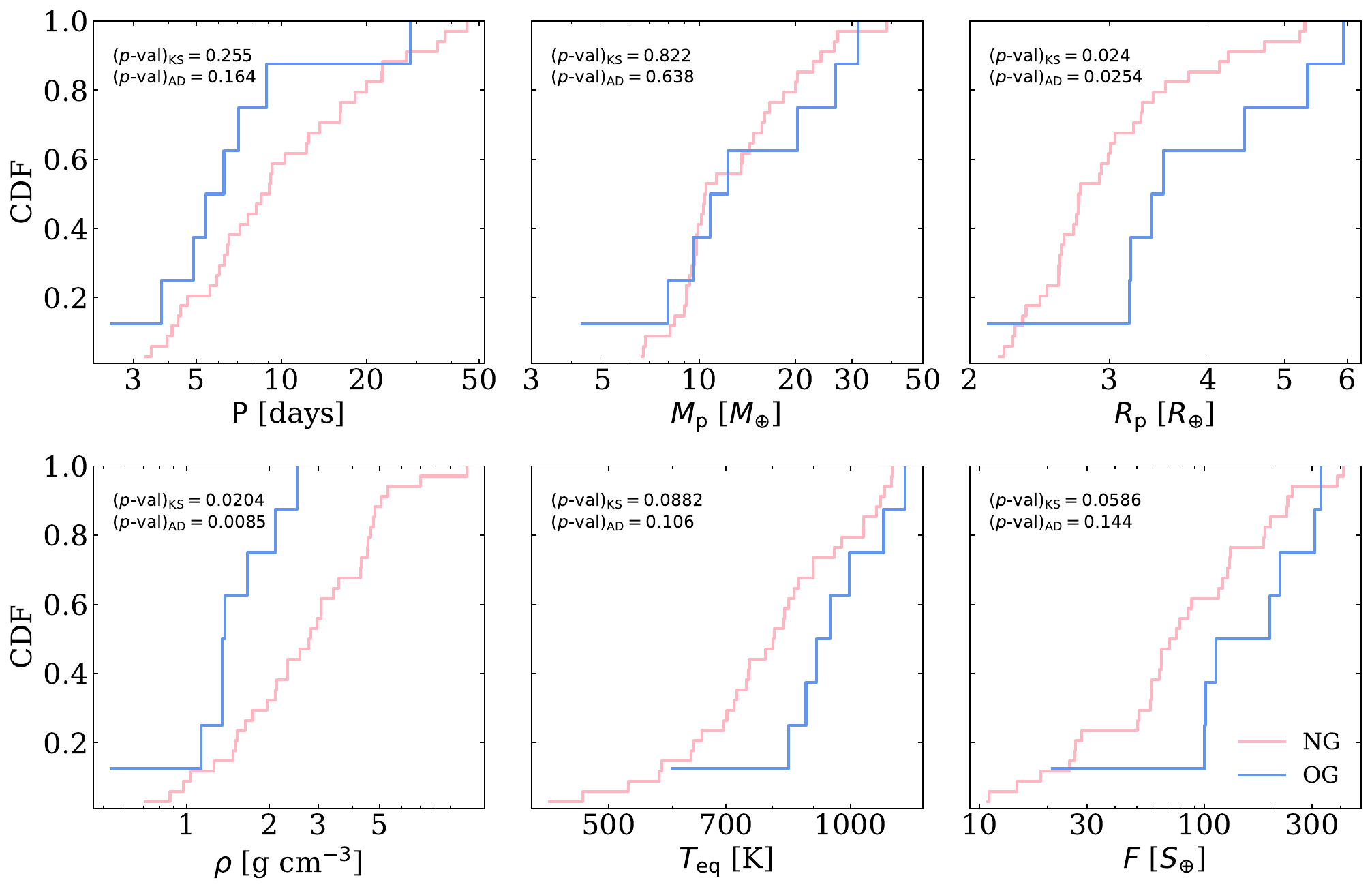}
   \caption{Empirical cumulative-distribution functions (CDFs) for planetary properties of inner (P < 50 d) Neptune-sized planets in systems with and without confirmed outer giants.
Left-to-right panels show orbital period, planet mass, planet radius (upper panel), planet density, equilibrium temperature and incident flux (bottom panel). Red curves correspond to NG systems, blue curves to OG systems. The inset labels give the two-sample KS and AD p-values for each parameter.
} 
       \label{fig:CDF_planets}%
    \end{figure*}

\begin{figure*}
   \centering
\includegraphics[width=0.8\textwidth] {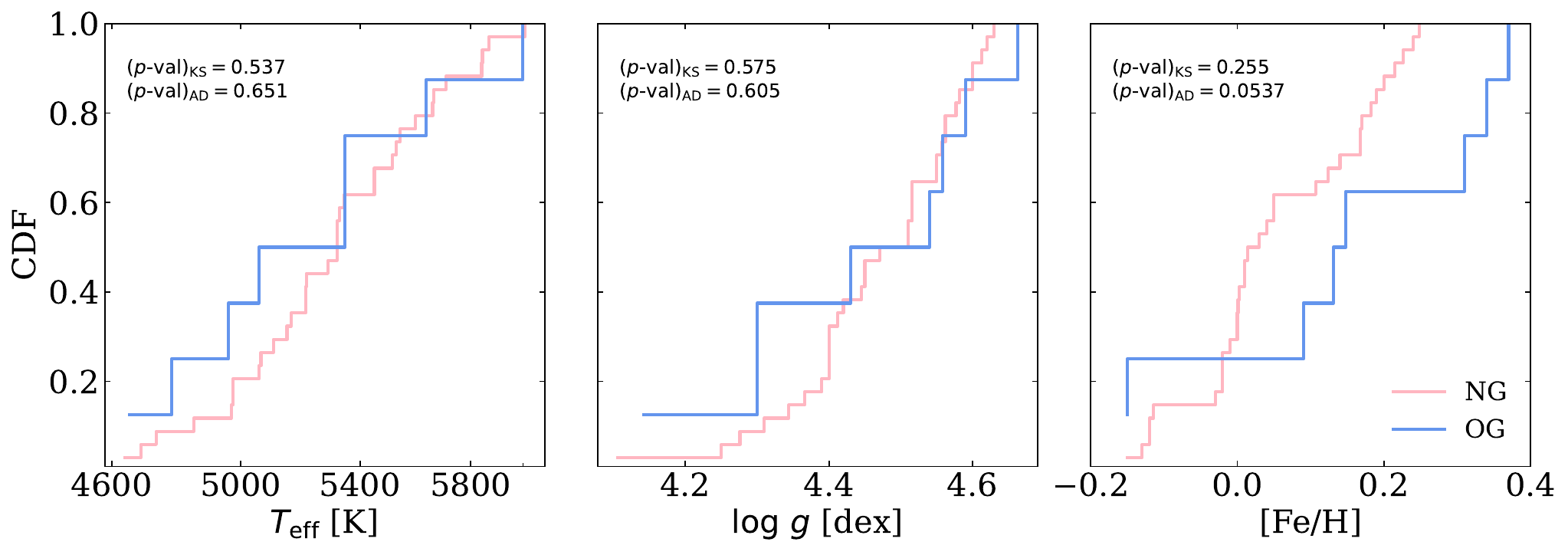}
   \caption{CDFs for host-star properties of the same planet sample as Fig~\ref{fig:CDF_planets}.
Panels show effective temperature ($T_{\rm eff}$), surface gravity ($\log g$); metallicity [Fe/H].
} 
   \label{fig:CDF_stars}%
    \end{figure*}

To quantify this observation further, and make sure it is not caused by any bias in planet or stellar properties, we show in Fig.~\ref{fig:CDF_planets} \& \ref{fig:CDF_stars}, the empirical cumulative distribution functions (CDFs) of the planets properties of orbital period $P$, mass $M_{\mathrm p}$, radius $R_{\mathrm p}$, density $\rho$, equilibrium temperature $T_{\rm eq}$ and incident flux $F$. Similarly, we give their stellar host properties of effective temperature $T_{\mathrm{eff}}$, surface gravity $\log g$ and metallicity [Fe/H]). 
We find the main parameter showing a strong difference between the OG and NG samples is the planet density, yielding a Kolmogorov–Smirnov (KS) p-value of $0.0204$ and an Anderson–Darling (AD) p-value of $0.0085$, suggesting that inner planets in OG systems may have somewhat lower bulk densities. This would be consistent with their radii being modestly inflated at a given mass. For the host-star, while spectral types are consistent between the samples, metallicity CDFs differ somewhat marginally (AD p-value $0.0506$), a level that is suggestive but not conclusive. Such a modest offset could simply reflect the well-established tendency for metal-rich stars to harbour giant planets \citep[e.g.,][]{FischerValenti05, Fulton21, Bryan24}, so metallicity may contribute to, but cannot by itself explain, the observed radius enhancement. We explore this dependence further in the following subsection. 

\subsection{Estimating the Radius-Inflation Offset in Outer-Giant Systems}
 
To estimate by how much the radii of inner planets in OG systems are larger compared to NG systems, we model both subsamples simultaneously with a single power law plus an additive offset in $\log R$, accounting for systems with outer giant planets
\begin{equation}
    \mathcal{R}_i = \alpha + \delta C_i + \beta \mathcal{M}_i
    \label{eq:mrmodel}
\end{equation}

where $\mathcal{R}_i\equiv \log{10}(R{_{\mathrm p},_i}/R_{\oplus})$ and $\mathcal{M}_i\equiv \log{10}(M_{{\mathrm p},i}/M_{\oplus})$ are the radius and mass of planet~$i$, $C_i$ is a binary indicator that equals `1' if planet $i$ orbits a star with at least one confirmed outer giant and `0' otherwise, $\alpha$ is the intercept of the baseline M--R relation, $\beta$ is its logarithmic slope, and $\delta$ captures the mean inflation suffered by planets in OG systems.

Given measurement uncertainties in both $\mathcal{M}$ and $\mathcal{R}$, we adopt a total least-squares likelihood following the formalism of \citet{Bashi17} and \citet{Muller24}, where the orthogonal residual of each point from Eq.(\ref{eq:mrmodel}) is divided by its propagated uncertainty in the same orthogonal direction, and explore parameter space with the \texttt{emcee} \citep{Foreman-Mackey13} MCMC sampler. The uninformative priors, as well as the resulting posterior medians, are summarised in Table~\ref{tab:fit_results} as well as plotted in Fig.~\ref{fig:MR} using a black dashed line to mark the best-fit linear trend with shaded purple colour to mark the added offset found in OG systems. The shaded grey regions mark the typical uncertainties in our fit.

While a simple linear fit is not intended to capture the full complexity of the M--R plane or the presence of multiple compositional families, we follow previous population-level studies \citep[e.g.,][]{WeissMarcy14, Bashi17, ChenKipping17, Edmondson23, Muller24} in adopting a shared-slope hierarchical model in this, somewhat modest phase-space region, as a pragmatic baseline, since our goal is to quantify a relative offset $\delta$ at fixed mass rather than to disentangle composition tracks.


\begin{table}
        \centering
        \caption{Prior and posterior distributions of model parameters fitting a total Least Squares model to the observed M--R relation}
        \label{tab:fit_results}
        \begin{tabular}{lcc} 
                \hline \hline
                Parameter & Prior & Posterior\\
                \hline
                $\alpha$ & $ \mathcal{U}(-1,1)$ & $-0.254_{-0.042}^{+0.040}$ \\
                $\delta$ & $\mathcal{U}(-1,1)$ & $0.068 \pm 0.014$\\
                $\beta$ & $\mathcal{U}(0,2)$&  $0.664_{-0.035}^{+0.037}$  \\
                \hline
        \end{tabular}
\end{table}

It is instructive to compare our empirical two-population M–R fit to established models. 
The power-law slope we find for the baseline population ($\beta \approx 0.66$) is in good agreement with recent global fits like the volatile rich population of \cite{Otegi20} who found a slope of $0.64 \pm 0.04$ and \cite{Muller24} who found a slope of $0.67 \pm 0.05$ in the $M_{\mathrm{p}}=5-120M_{\mathrm{\oplus}}$ mass range. 

In linear units, we find the radius offset $\delta = 0.068 \pm 0.014$ corresponds to an average radius inflation factor of $16.9\pm 3.8 \,\%$, i.e. approximately $17\%$ larger radii for inner planets in systems with an OG compared to similar-mass planets in NG systems. The inflated radius in OG systems is significant at the $\approx4\sigma$ level and suggests that OG companions influence the envelope properties or subsequent evolution of inner planets, either by facilitating more efficient envelope accretion during formation or by mitigating post-formation atmospheric loss processes.

\subsection{Consistency Check with Homogeneous Parameters}

To quantify the impact of catalogue heterogeneity, we constructed a homogeneous control set of host and planet parameters. We first adopted uniform stellar atmospheric properties $T_{\mathrm{eff}}$, $\log g$ and [Fe/H]) using the derived \cite{Andrae23} \gaia XP spectra, then inferred stellar mass $M_\star^{\mathrm{XP}}$ and radius $R_\star^{\mathrm{XP}}$ properties with the empirical calibrations of \cite{Torres10}. Planetary quantities were propagated to these revised hosts using
$R_{p}^{\mathrm{XP}} = R_{p}\,\Big(R_\star^{\mathrm{XP}} / {R_\star}\Big)$, 
$ M_{p}^{\mathrm{XP}} = M_{p}\,\Big(M_\star^{\mathrm{XP}}/ M_\star\Big)^{2/3}$.
We then ran the hierarchical M--R fit. The architecture offset changed from $\delta=0.068\pm0.014$ to $\delta_{\rm XP}=0.053\pm0.015$ in the homogeneous set, while the shared slope and intrinsic scatter remained consistent within $1\sigma$. Thus, enforcing homogeneity shifts the offset by only $\sim0.015$ dex, and it remains significant ($\sim\!3.5\sigma$), indicating that moderate parameter inhomogeneity does not drive our demographic result.

\subsection{Testing for Metallicity Biases}

Host star metallicity is known to correlate with planetary properties. Using \textit{Gaia} DR3 spectroscopic catalogue, \cite{deLaverny25} showed that the median host metallicity rises smoothly with planet radius for sub-Jovian planets (their Fig. 21), indicating a continuous dependence of typical planet size on [Fe/H]. In the hot-Neptune regime specifically, \cite{VissapragadaBehmard25} find that the hottest Neptunes preferentially orbit metal-rich stars. These trends motivate a direct test of whether the offset we measure between OG and NG systems could be driven primarily by metallicity.

To assess whether these metallicity trends could explain our architecture-dependent offset, we repeated our hierarchical M--R fit using the same likelihood and priors, but with the binary indicator $C_i$ defined by host metallicity rather than system architecture. Specifically, we divided the sample at [Fe/H] = 0 and estimated the radius shift parameter $\delta_{\rm [Fe/H]}$.
We find a $\delta_{\rm [Fe/H]} = -0.012 \pm 0.015$, statistically consistent with zero radius offset between metal-poor and metal-rich hosts at fixed mass. Compared to our architecture-based split, this result differs by $\sim 0.08$ dex, indicating that metallicity alone cannot account for the OG–NG inflation signal we report.

\section{Discussion}
\label{sec:Discussion}
\subsection{An Architecture-Dependent M--R Relation}

Our results can be thought of as introducing an `architecture-dependent' tweak to standard M--R relations. Similar to studies that attempt to include additional parameters beyond mass, for instance, heavy elements fraction, incident flux or equilibrium temperature \citep{Enoch12, Ulmer-Moll19, Edmondson23}, to explain scatter in the M–R relation, our findings suggest that a binary parameter for outer giant presence could be another such factor. Conceptually, one could incorporate it as separate M–R relations or an additive term in predictive models of planet radius. 

The new architecture term dovetails with the `peas-in-a-pod' pattern: planets in a compact chain remain nearly uniform in size and grow slightly larger with orbital distance \citep{Weiss18, MillhollandWinn21}. Our outer-giant offset effectively raises that entire size ladder without changing the relative step between neighbours, implying a system-wide boost.


\subsection{Physical Mechanisms Driving Radius Inflation}

Several formation mechanisms could naturally produce larger radii for inner sub-Neptunes in systems with outer giants. Hydrodynamic simulations of planet–disc interactions show that a Jupiter-mass planet can modify disc dispersal: the gap opened by the giant does not isolate the inner disc as previously assumed but instead is partly refilled by wind-driven flows \citep{Weber24}. This process boosts mass transport across the gap and leaves a denser inner disc that persists longer. Prolonged access to nebular gas enables close-in sub-Neptunes to accrete or retain more substantial H/He envelopes than they otherwise would, contributing to inflated planet sizes. 

Another promising explanation is the volatile enrichment of the inner planets. In pebble-accretion models, a growing giant core can intercept the flow of icy pebbles and release a surge of water vapour interior to its orbit if it forms inside the ice line \citep{Bitsch21}. In addition, a cold Jupiter can excite icy planetesimals, increase water delivery, and thereby lower mean density even without extra H/He accretion \citep{OBrien14}. Inner sub-Neptunes that accrete gas under these conditions would incorporate significant water content (e.g. steam atmospheres or higher water fraction in their envelopes), making them lower-density and thus physically larger for a given mass. 

\subsection{Limitations, Biases, and Future Tests}
While we outlined a few physical mechanisms to explain the systematically larger radii of close-in sub-Neptunes in OG systems, it is still important to be cautious with our results, as the current OG subsample (8 planets) is small and the statistical power is limited. Future high-precision masses from ESPRESSO \citep{ESPRESSO21} and radii from PLATO \citep{PLATO25} will more than double the OG sample, enabling a hierarchical treatment of detection completeness. Furthermore, planetary radii evolve for hundreds of Myr as interiors cool and atmospheres escape; younger sub-Neptunes are systematically `puffier,' while older ones contract or lose H/He envelopes \citep{Berger20, Rogers25}. If the OG and NG hosts differ in age, then secular evolution could mimic or mask the radius enhancement we report. In that sense, the PLATO mission consortium is expected to provide the community with a 10\% accuracy \citep{PLATO25} in stellar age to many planet host stars, thereby allowing us to test this hypothesis.

Another potential confounder is tidally driven inflation in misaligned (high-obliquity) planets.  Obliquity tides can inject $\gtrsim 10^{-5} L_\star$ of heat into the envelope \citep{Millholland19, SethiMillholland25}, enlarging radii by $10-30 \%$. Although only one misaligned planet currently appears in the OG subsample (out of $8$), secular perturbations from an outer giant are expected to raise stellar obliquities over Gyr timescales \citep{LinOgilvie17}.  Distinguishing whether radius inflation arises from (i) extra gas accretion in the proto-planetary disc or (ii) ongoing tidal heating will therefore require a larger set of spin–orbit angle measurements, something that the upcoming surveys should provide.

Finally, analysis of archival datasets, such as the Kepler Giant Planet Survey \citep[KGPS][]{Weiss24}, where RV trends have been observed in some systems, can also extend our analysis to include suspected OG systems. A careful analysis of all RV time-series of the planets considered in this work is beyond the scope of this study and should be left for future work \citep[e.g.,][]{Osborne25}. Nevertheless, independent occurrence-rate studies of giant planets in systems with close-in small (CS) planets \citep[e.g.,][]{VanZandt25} have recently reported a moderate enhancement of distant giants (DG), with $P({\rm DG|CS}) = 31^{+11}_{-12}\%$. Our current sample, with 8 OG systems versus 34 NG systems, is broadly consistent with this expectation. This suggests that our OG fraction is not strongly incomplete, further supporting the robustness of our demographic comparison.

\section{Summary}
\label{sec:Summary}
In this paper, we compiled a uniformly selected sample of close-in ($P < 50 $ days) Neptune-sized planets and found that in systems hosting a confirmed outer giant ($P=300-10,000$ days, $M_p \geq 80 M_{\oplus}$) lie systematically farther above the 100\% H$_2$O M--R line than those in giant-free systems. In addition, using a total-least-squares MCMC fit to the M--R relation, we were able to estimate that these inner planets in OG systems are, on average, $16.9 \pm 3.8\,\%$ larger at fixed mass than those in NG systems, a $\approx4\sigma$ result independent of stellar or orbital-period differences.


The architecture-sensitive M--R relation we report opens a new diagnostic window on planet formation. 
 If the radius enhancement persists, given our limited sample, it will provide a stringent benchmark for population-synthesis models that seek to reproduce the intertwined emergence of cold Jupiters and close-in sub-Neptunes.

\section*{Acknowledgements}
We are grateful to the anonymous referee, whose careful review and insightful suggestions strengthened our analysis. We wish to thank Didier Queloz and Ravit Helled for helpful discussions and suggestions. DB acknowledges the support of the Blavatnik family and the British Friends of the Hebrew University (BFHU) as part of the Blavatnik Cambridge Fellowship. This research has made use of the NASA Exoplanet Archive, which is operated by the California Institute of Technology, under contract with the National Aeronautics and Space Administration under the Exoplanet Exploration Program. This work has made use of data from the European Space Agency (ESA) mission {\it Gaia} (\url{https://www.cosmos.esa.int/gaia}), processed by the {\it Gaia}
Data Processing and Analysis Consortium (DPAC,
\url{https://www.cosmos.esa.int/web/gaia/dpac/consortium}). Funding for the DPAC
has been provided by national institutions, in particular the institutions
participating in the {\it Gaia} Multilateral Agreement. This research also made use of TOPCAT \citep{Taylor05}, an interactive
graphical viewer and editor for tabular data. This work made use of Astropy,
a community-developed core Python package and an
ecosystem of tools and resources for astronomy \citep{Astropy22}.

\section*{Data Availability}

The research presented in this article primarily relies on publicly available and accessible data from the NASA Exoplanet Archive.



\bibliographystyle{mnras}
\bibliography{example} 

\begin{thebibliography}{}
\makeatletter
\relax
\def\mn@urlcharsother{\let\do\@makeother \do\$\do\&\do\#\do\^\do\_\do\%\do\~}
\def\mn@doi{\begingroup\mn@urlcharsother \@ifnextchar [ {\mn@doi@} {\mn@doi@[]}}
\def\mn@doi@[#1]#2{\def\@tempa{#1}\ifx\@tempa\@empty \href {http://dx.doi.org/#2} {doi:#2}\else \href {http://dx.doi.org/#2} {#1}\fi \endgroup}
\def\mn@eprint#1#2{\mn@eprint@#1:#2::\@nil}
\def\mn@eprint@arXiv#1{\href {http://arxiv.org/abs/#1} {{\tt arXiv:#1}}}
\def\mn@eprint@dblp#1{\href {http://dblp.uni-trier.de/rec/bibtex/#1.xml} {dblp:#1}}
\def\mn@eprint@#1:#2:#3:#4\@nil{\def\@tempa {#1}\def\@tempb {#2}\def\@tempc {#3}\ifx \@tempc \@empty \let \@tempc \@tempb \let \@tempb \@tempa \fi \ifx \@tempb \@empty \def\@tempb {arXiv}\fi \@ifundefined {mn@eprint@\@tempb}{\@tempb:\@tempc}{\expandafter \expandafter \csname mn@eprint@\@tempb\endcsname \expandafter{\@tempc}}}

\bibitem[\protect\citeauthoryear{{Andrae}, {Rix}  \& {Chandra}}{{Andrae} et~al.}{2023}]{Andrae23}
{Andrae} R.,  {Rix} H.-W.,   {Chandra} V.,  2023, \mn@doi [\apjs] {10.3847/1538-4365/acd53e}, \href {https://ui.adsabs.harvard.edu/abs/2023ApJS..267....8A} {267, 8}

\bibitem[\protect\citeauthoryear{{Astropy Collaboration} et~al.,}{{Astropy Collaboration} et~al.}{2022}]{Astropy22}
{Astropy Collaboration} et~al., 2022, \mn@doi [\apj] {10.3847/1538-4357/ac7c74}, \href {https://ui.adsabs.harvard.edu/abs/2022ApJ...935..167A} {935, 167}

\bibitem[\protect\citeauthoryear{{Bashi} \& {Zucker}}{{Bashi} \& {Zucker}}{2021}]{BashiZucker21}
{Bashi} D.,  {Zucker} S.,  2021, \mn@doi [\aap] {10.1051/0004-6361/202140699}, \href {https://ui.adsabs.harvard.edu/abs/2021A&A...651A..61B} {651, A61}

\bibitem[\protect\citeauthoryear{{Bashi}, {Helled}, {Zucker}  \& {Mordasini}}{{Bashi} et~al.}{2017}]{Bashi17}
{Bashi} D.,  {Helled} R.,  {Zucker} S.,   {Mordasini} C.,  2017, \mn@doi [\aap] {10.1051/0004-6361/201629922}, \href {https://ui.adsabs.harvard.edu/abs/2017A&A...604A..83B} {604, A83}

\bibitem[\protect\citeauthoryear{{Bashi}, {Mazeh}  \& {Faigler}}{{Bashi} et~al.}{2024}]{Bashi24}
{Bashi} D.,  {Mazeh} T.,   {Faigler} S.,  2024, \mn@doi [\aj] {10.3847/1538-3881/ad5ffa}, \href {https://ui.adsabs.harvard.edu/abs/2024AJ....168..115B} {168, 115}

\bibitem[\protect\citeauthoryear{{Berger}, {Huber}, {Gaidos}, {van Saders}  \& {Weiss}}{{Berger} et~al.}{2020}]{Berger20}
{Berger} T.~A.,  {Huber} D.,  {Gaidos} E.,  {van Saders} J.~L.,   {Weiss} L.~M.,  2020, \mn@doi [\aj] {10.3847/1538-3881/aba18a}, \href {https://ui.adsabs.harvard.edu/abs/2020AJ....160..108B} {160, 108}

\bibitem[\protect\citeauthoryear{{Best}, {Sefilian}  \& {Petrovich}}{{Best} et~al.}{2024}]{Best24}
{Best} M.,  {Sefilian} A.~A.,   {Petrovich} C.,  2024, \mn@doi [\apj] {10.3847/1538-4357/ad0965}, \href {https://ui.adsabs.harvard.edu/abs/2024ApJ...960...89B} {960, 89}

\bibitem[\protect\citeauthoryear{{Bitsch} \& {Izidoro}}{{Bitsch} \& {Izidoro}}{2023}]{BitschIzidoro23}
{Bitsch} B.,  {Izidoro} A.,  2023, \mn@doi [\aap] {10.1051/0004-6361/202245040}, \href {https://ui.adsabs.harvard.edu/abs/2023A&A...674A.178B} {674, A178}

\bibitem[\protect\citeauthoryear{{Bitsch}, {Raymond}, {Buchhave}, {Bello-Arufe}, {Rathcke}  \& {Schneider}}{{Bitsch} et~al.}{2021}]{Bitsch21}
{Bitsch} B.,  {Raymond} S.~N.,  {Buchhave} L.~A.,  {Bello-Arufe} A.,  {Rathcke} A.~D.,   {Schneider} A.~D.,  2021, \mn@doi [\aap] {10.1051/0004-6361/202140793}, \href {https://ui.adsabs.harvard.edu/abs/2021A&A...649L...5B} {649, L5}

\bibitem[\protect\citeauthoryear{{Bonomo} et~al.,}{{Bonomo} et~al.}{2023}]{Bonomo23}
{Bonomo} A.~S.,  et~al., 2023, \mn@doi [\aap] {10.1051/0004-6361/202346211}, \href {https://ui.adsabs.harvard.edu/abs/2023A&A...677A..33B} {677, A33}

\bibitem[\protect\citeauthoryear{{Bryan} \& {Lee}}{{Bryan} \& {Lee}}{2024}]{Bryan24}
{Bryan} M.~L.,  {Lee} E.~J.,  2024, \mn@doi [\apjl] {10.3847/2041-8213/ad5013}, \href {https://ui.adsabs.harvard.edu/abs/2024ApJ...968L..25B} {968, L25}

\bibitem[\protect\citeauthoryear{{Bryant}, {Bayliss}  \& {Van Eylen}}{{Bryant} et~al.}{2023}]{Bryant23}
{Bryant} E.~M.,  {Bayliss} D.,   {Van Eylen} V.,  2023, \mn@doi [\mnras] {10.1093/mnras/stad626}, \href {https://ui.adsabs.harvard.edu/abs/2023MNRAS.521.3663B} {521, 3663}

\bibitem[\protect\citeauthoryear{{Chen} \& {Kipping}}{{Chen} \& {Kipping}}{2017}]{ChenKipping17}
{Chen} J.,  {Kipping} D.,  2017, \mn@doi [\apj] {10.3847/1538-4357/834/1/17}, \href {https://ui.adsabs.harvard.edu/abs/2017ApJ...834...17C} {834, 17}

\bibitem[\protect\citeauthoryear{{Christiansen} et~al.,}{{Christiansen} et~al.}{2025}]{NASA_Archive}
{Christiansen} J.~L.,  et~al., 2025, \mn@doi [arXiv e-prints] {10.48550/arXiv.2506.03299}, \href {https://ui.adsabs.harvard.edu/abs/2025arXiv250603299C} {p. arXiv:2506.03299}

\bibitem[\protect\citeauthoryear{{Edmondson}, {Norris}  \& {Kerins}}{{Edmondson} et~al.}{2023}]{Edmondson23}
{Edmondson} K.,  {Norris} J.,   {Kerins} E.,  2023, \mn@doi [arXiv e-prints] {10.48550/arXiv.2310.16733}, \href {https://ui.adsabs.harvard.edu/abs/2023arXiv231016733E} {p. arXiv:2310.16733}

\bibitem[\protect\citeauthoryear{{Emsenhuber}, {Mordasini}  \& {Burn}}{{Emsenhuber} et~al.}{2023}]{Emsenhuber23}
{Emsenhuber} A.,  {Mordasini} C.,   {Burn} R.,  2023, \mn@doi [European Physical Journal Plus] {10.1140/epjp/s13360-023-03784-x}, \href {https://ui.adsabs.harvard.edu/abs/2023EPJP..138..181E} {138, 181}

\bibitem[\protect\citeauthoryear{{Enoch}, {Collier Cameron}  \& {Horne}}{{Enoch} et~al.}{2012}]{Enoch12}
{Enoch} B.,  {Collier Cameron} A.,   {Horne} K.,  2012, \mn@doi [\aap] {10.1051/0004-6361/201117317}, \href {https://ui.adsabs.harvard.edu/abs/2012A&A...540A..99E} {540, A99}

\bibitem[\protect\citeauthoryear{{Fischer} \& {Valenti}}{{Fischer} \& {Valenti}}{2005}]{FischerValenti05}
{Fischer} D.~A.,  {Valenti} J.,  2005, \mn@doi [\apj] {10.1086/428383}, \href {https://ui.adsabs.harvard.edu/abs/2005ApJ...622.1102F} {622, 1102}

\bibitem[\protect\citeauthoryear{{Foreman-Mackey} et~al.,}{{Foreman-Mackey} et~al.}{2013}]{Foreman-Mackey13}
{Foreman-Mackey} D.,  et~al., 2013, {emcee: The MCMC Hammer}, Astrophysics Source Code Library, record ascl:1303.002 (\mn@eprint {ascl} {1303.002})

\bibitem[\protect\citeauthoryear{{Fulton} et~al.,}{{Fulton} et~al.}{2017}]{Fulton17}
{Fulton} B.~J.,  et~al., 2017, \mn@doi [\aj] {10.3847/1538-3881/aa80eb}, \href {https://ui.adsabs.harvard.edu/abs/2017AJ....154..109F} {154, 109}

\bibitem[\protect\citeauthoryear{{Fulton} et~al.,}{{Fulton} et~al.}{2021}]{Fulton21}
{Fulton} B.~J.,  et~al., 2021, \mn@doi [\apjs] {10.3847/1538-4365/abfcc1}, \href {https://ui.adsabs.harvard.edu/abs/2021ApJS..255...14F} {255, 14}

\bibitem[\protect\citeauthoryear{{Gaia Collaboration} et~al.,}{{Gaia Collaboration} et~al.}{2023}]{GaiaDR3}
{Gaia Collaboration} et~al., 2023, \mn@doi [\aap] {10.1051/0004-6361/202243940}, \href {https://ui.adsabs.harvard.edu/abs/2023A&A...674A...1G} {674, A1}

\bibitem[\protect\citeauthoryear{{Gilbert} \& {Fabrycky}}{{Gilbert} \& {Fabrycky}}{2020}]{GilbertFabrycky20}
{Gilbert} G.~J.,  {Fabrycky} D.~C.,  2020, \mn@doi [\aj] {10.3847/1538-3881/ab8e3c}, \href {https://ui.adsabs.harvard.edu/abs/2020AJ....159..281G} {159, 281}

\bibitem[\protect\citeauthoryear{{Han}, {Robertson}, {Brandt}, {Kanodia}, {Ca{\~n}as}, {Shporer}, {Ricker}  \& {Beard}}{{Han} et~al.}{2025}]{Han25}
{Han} T.,  {Robertson} P.,  {Brandt} T.~D.,  {Kanodia} S.,  {Ca{\~n}as} C.,  {Shporer} A.,  {Ricker} G.,   {Beard} C.,  2025, \mn@doi [\apjl] {10.3847/2041-8213/ade794}, \href {https://ui.adsabs.harvard.edu/abs/2025ApJ...988L...4H} {988, L4}

\bibitem[\protect\citeauthoryear{{He} \& {Weiss}}{{He} \& {Weiss}}{2023}]{HeWeiss23}
{He} M.~Y.,  {Weiss} L.~M.,  2023, \mn@doi [\aj] {10.3847/1538-3881/acdd56}, \href {https://ui.adsabs.harvard.edu/abs/2023AJ....166...36H} {166, 36}

\bibitem[\protect\citeauthoryear{{Ho} \& {Van Eylen}}{{Ho} \& {Van Eylen}}{2023}]{HoVanEylen23}
{Ho} C. S.~K.,  {Van Eylen} V.,  2023, \mn@doi [\mnras] {10.1093/mnras/stac3802}, \href {https://ui.adsabs.harvard.edu/abs/2023MNRAS.519.4056H} {519, 4056}

\bibitem[\protect\citeauthoryear{{Kane} \& {Wittenmyer}}{{Kane} \& {Wittenmyer}}{2024}]{Kane24}
{Kane} S.~R.,  {Wittenmyer} R.~A.,  2024, \mn@doi [\apjl] {10.3847/2041-8213/ad2463}, \href {https://ui.adsabs.harvard.edu/abs/2024ApJ...962L..21K} {962, L21}

\bibitem[\protect\citeauthoryear{{Lin} \& {Ogilvie}}{{Lin} \& {Ogilvie}}{2017}]{LinOgilvie17}
{Lin} Y.,  {Ogilvie} G.~I.,  2017, \mn@doi [\mnras] {10.1093/mnras/stx540}, \href {https://ui.adsabs.harvard.edu/abs/2017MNRAS.468.1387L} {468, 1387}

\bibitem[\protect\citeauthoryear{{Lindegren} et~al.,}{{Lindegren} et~al.}{2021}]{Lindegren21}
{Lindegren} L.,  et~al., 2021, \mn@doi [\aap] {10.1051/0004-6361/202039709}, \href {https://ui.adsabs.harvard.edu/abs/2021A&A...649A...2L} {649, A2}

\bibitem[\protect\citeauthoryear{{Lozovsky} \& {Perets}}{{Lozovsky} \& {Perets}}{2025}]{LozovskyPerets25}
{Lozovsky} M.,  {Perets} H.~B.,  2025, \mn@doi [arXiv e-prints] {10.48550/arXiv.2508.13274}, \href {https://ui.adsabs.harvard.edu/abs/2025arXiv250813274L} {p. arXiv:2508.13274}

\bibitem[\protect\citeauthoryear{{Millholland}}{{Millholland}}{2019}]{Millholland19}
{Millholland} S.,  2019, \mn@doi [\apj] {10.3847/1538-4357/ab4c3f}, \href {https://ui.adsabs.harvard.edu/abs/2019ApJ...886...72M} {886, 72}

\bibitem[\protect\citeauthoryear{{Millholland} \& {Winn}}{{Millholland} \& {Winn}}{2021}]{MillhollandWinn21}
{Millholland} S.~C.,  {Winn} J.~N.,  2021, \mn@doi [\apjl] {10.3847/2041-8213/ac2c77}, \href {https://ui.adsabs.harvard.edu/abs/2021ApJ...920L..34M} {920, L34}

\bibitem[\protect\citeauthoryear{{Mills} \& {Mazeh}}{{Mills} \& {Mazeh}}{2017}]{MillsMazeh17}
{Mills} S.~M.,  {Mazeh} T.,  2017, \mn@doi [\apjl] {10.3847/2041-8213/aa67eb}, \href {https://ui.adsabs.harvard.edu/abs/2017ApJ...839L...8M} {839, L8}

\bibitem[\protect\citeauthoryear{{M{\"u}ller}, {Baron}, {Helled}, {Bouchy}  \& {Parc}}{{M{\"u}ller} et~al.}{2024}]{Muller24}
{M{\"u}ller} S.,  {Baron} J.,  {Helled} R.,  {Bouchy} F.,   {Parc} L.,  2024, \mn@doi [\aap] {10.1051/0004-6361/202348690}, \href {https://ui.adsabs.harvard.edu/abs/2024A&A...686A.296M} {686, A296}

\bibitem[\protect\citeauthoryear{{O'Brien}, {Walsh}, {Morbidelli}, {Raymond}  \& {Mandell}}{{O'Brien} et~al.}{2014}]{OBrien14}
{O'Brien} D.~P.,  {Walsh} K.~J.,  {Morbidelli} A.,  {Raymond} S.~N.,   {Mandell} A.~M.,  2014, \mn@doi [\icarus] {10.1016/j.icarus.2014.05.009}, \href {https://ui.adsabs.harvard.edu/abs/2014Icar..239...74O} {239, 74}

\bibitem[\protect\citeauthoryear{{Osborne}, {Nielsen}, {Van Eylen}  \& {Barrag{\'a}n}}{{Osborne} et~al.}{2025}]{Osborne25}
{Osborne} H.~L.~M.,  {Nielsen} L.~D.,  {Van Eylen} V.,   {Barrag{\'a}n} O.,  2025, \mn@doi [\aap] {10.1051/0004-6361/202452127}, \href {https://ui.adsabs.harvard.edu/abs/2025A&A...693A...4O} {693, A4}

\bibitem[\protect\citeauthoryear{{Otegi}, {Bouchy}  \& {Helled}}{{Otegi} et~al.}{2020}]{Otegi20}
{Otegi} J.~F.,  {Bouchy} F.,   {Helled} R.,  2020, \mn@doi [\aap] {10.1051/0004-6361/201936482}, \href {https://ui.adsabs.harvard.edu/abs/2020A&A...634A..43O} {634, A43}

\bibitem[\protect\citeauthoryear{{Pepe} et~al.,}{{Pepe} et~al.}{2021}]{ESPRESSO21}
{Pepe} F.,  et~al., 2021, \mn@doi [\aap] {10.1051/0004-6361/202038306}, \href {https://ui.adsabs.harvard.edu/abs/2021A&A...645A..96P} {645, A96}

\bibitem[\protect\citeauthoryear{{Rauer} et~al.,}{{Rauer} et~al.}{2025}]{PLATO25}
{Rauer} H.,  et~al., 2025, \mn@doi [Experimental Astronomy] {10.1007/s10686-025-09985-9}, \href {https://ui.adsabs.harvard.edu/abs/2025ExA....59...26R} {59, 26}

\bibitem[\protect\citeauthoryear{{Rice}, {Steffen}  \& {Vazan}}{{Rice} et~al.}{2024}]{Rice24}
{Rice} D.~R.,  {Steffen} J.~H.,   {Vazan} A.,  2024, \mn@doi [\apjl] {10.3847/2041-8213/ad73db}, \href {https://ui.adsabs.harvard.edu/abs/2024ApJ...973L...4R} {973, L4}

\bibitem[\protect\citeauthoryear{{Rogers}}{{Rogers}}{2025}]{Rogers25}
{Rogers} J.~G.,  2025, \mn@doi [\mnras] {10.1093/mnras/staf628}, \href {https://ui.adsabs.harvard.edu/abs/2025MNRAS.539.2230R} {539, 2230}

\bibitem[\protect\citeauthoryear{{Rosenthal} et~al.,}{{Rosenthal} et~al.}{2022}]{Rosenthal22}
{Rosenthal} L.~J.,  et~al., 2022, \mn@doi [\apjs] {10.3847/1538-4365/ac7230}, \href {https://ui.adsabs.harvard.edu/abs/2022ApJS..262....1R} {262, 1}

\bibitem[\protect\citeauthoryear{{Sabotta} et~al.,}{{Sabotta} et~al.}{2021}]{Sabotta21}
{Sabotta} S.,  et~al., 2021, \mn@doi [\aap] {10.1051/0004-6361/202140968}, \href {https://ui.adsabs.harvard.edu/abs/2021A&A...653A.114S} {653, A114}

\bibitem[\protect\citeauthoryear{{Sethi} \& {Millholland}}{{Sethi} \& {Millholland}}{2025}]{SethiMillholland25}
{Sethi} R.,  {Millholland} S.~C.,  2025, \mn@doi [\apj] {10.3847/1538-4357/ade883}, \href {https://ui.adsabs.harvard.edu/abs/2025ApJ...988..247S} {988, 247}

\bibitem[\protect\citeauthoryear{{Sobski} \& {Millholland}}{{Sobski} \& {Millholland}}{2023}]{Sobski23}
{Sobski} N.,  {Millholland} S.~C.,  2023, \mn@doi [\apj] {10.3847/1538-4357/ace966}, \href {https://ui.adsabs.harvard.edu/abs/2023ApJ...954..137S} {954, 137}

\bibitem[\protect\citeauthoryear{{Taylor}}{{Taylor}}{2005}]{Taylor05}
{Taylor} M.~B.,  2005, in {Shopbell} P.,  {Britton} M.,   {Ebert} R.,  eds,  Astronomical Society of the Pacific Conference Series Vol. 347, Astronomical Data Analysis Software and Systems XIV. p.~29

\bibitem[\protect\citeauthoryear{{Torres}, {Andersen}  \& {Gim{\'e}nez}}{{Torres} et~al.}{2010}]{Torres10}
{Torres} G.,  {Andersen} J.,   {Gim{\'e}nez} A.,  2010, \mn@doi [\aapr] {10.1007/s00159-009-0025-1}, \href {https://ui.adsabs.harvard.edu/abs/2010A&ARv..18...67T} {18, 67}

\bibitem[\protect\citeauthoryear{{Ulmer-Moll}, {Santos}, {Figueira}, {Brinchmann}  \& {Faria}}{{Ulmer-Moll} et~al.}{2019}]{Ulmer-Moll19}
{Ulmer-Moll} S.,  {Santos} N.~C.,  {Figueira} P.,  {Brinchmann} J.,   {Faria} J.~P.,  2019, \mn@doi [\aap] {10.1051/0004-6361/201936049}, \href {https://ui.adsabs.harvard.edu/abs/2019A&A...630A.135U} {630, A135}

\bibitem[\protect\citeauthoryear{{Van Eylen}, {Agentoft}, {Lundkvist}, {Kjeldsen}, {Owen}, {Fulton}, {Petigura}  \& {Snellen}}{{Van Eylen} et~al.}{2018}]{VanEylen18}
{Van Eylen} V.,  {Agentoft} C.,  {Lundkvist} M.~S.,  {Kjeldsen} H.,  {Owen} J.~E.,  {Fulton} B.~J.,  {Petigura} E.,   {Snellen} I.,  2018, \mn@doi [\mnras] {10.1093/mnras/sty1783}, \href {https://ui.adsabs.harvard.edu/abs/2018MNRAS.479.4786V} {479, 4786}

\bibitem[\protect\citeauthoryear{{Van Zandt} et~al.,}{{Van Zandt} et~al.}{2025}]{VanZandt25}
{Van Zandt} J.,  et~al., 2025, \mn@doi [\aj] {10.3847/1538-3881/adbbed}, \href {https://ui.adsabs.harvard.edu/abs/2025AJ....169..235V} {169, 235}

\bibitem[\protect\citeauthoryear{{Vissapragada} \& {Behmard}}{{Vissapragada} \& {Behmard}}{2025}]{VissapragadaBehmard25}
{Vissapragada} S.,  {Behmard} A.,  2025, \mn@doi [\aj] {10.3847/1538-3881/ada143}, \href {https://ui.adsabs.harvard.edu/abs/2025AJ....169..117V} {169, 117}

\bibitem[\protect\citeauthoryear{{Weber}, {Picogna}  \& {Ercolano}}{{Weber} et~al.}{2024}]{Weber24}
{Weber} M.~L.,  {Picogna} G.,   {Ercolano} B.,  2024, \mn@doi [\aap] {10.1051/0004-6361/202348596}, \href {https://ui.adsabs.harvard.edu/abs/2024A&A...686A..53W} {686, A53}

\bibitem[\protect\citeauthoryear{{Weiss} \& {Marcy}}{{Weiss} \& {Marcy}}{2014}]{WeissMarcy14}
{Weiss} L.~M.,  {Marcy} G.~W.,  2014, \mn@doi [\apjl] {10.1088/2041-8205/783/1/L6}, \href {https://ui.adsabs.harvard.edu/abs/2014ApJ...783L...6W} {783, L6}

\bibitem[\protect\citeauthoryear{{Weiss} et~al.,}{{Weiss} et~al.}{2018}]{Weiss18}
{Weiss} L.~M.,  et~al., 2018, \mn@doi [\aj] {10.3847/1538-3881/aa9ff6}, \href {https://ui.adsabs.harvard.edu/abs/2018AJ....155...48W} {155, 48}

\bibitem[\protect\citeauthoryear{{Weiss} et~al.,}{{Weiss} et~al.}{2024}]{Weiss24}
{Weiss} L.~M.,  et~al., 2024, \mn@doi [\apjs] {10.3847/1538-4365/ad0cab}, \href {https://ui.adsabs.harvard.edu/abs/2024ApJS..270....8W} {270, 8}

\bibitem[\protect\citeauthoryear{{Zeng} et~al.,}{{Zeng} et~al.}{2019}]{Zeng2019}
{Zeng} L.,  et~al., 2019, \mn@doi [Proceedings of the National Academy of Science] {10.1073/pnas.1812905116}, \href {https://ui.adsabs.harvard.edu/abs/2019PNAS..116.9723Z} {116, 9723}

\bibitem[\protect\citeauthoryear{{de Laverny}, {Ligi}, {Crida}, {Recio-Blanco}  \& {Palicio}}{{de Laverny} et~al.}{2025}]{deLaverny25}
{de Laverny} P.,  {Ligi} R.,  {Crida} A.,  {Recio-Blanco} A.,   {Palicio} P.~A.,  2025, \mn@doi [\aap] {10.1051/0004-6361/202554739}, \href {https://ui.adsabs.harvard.edu/abs/2025A&A...699A.100D} {699, A100}

\makeatother
\end{thebibliography}


\appendix

\section{Inner-planet M--R sample}
\label{app:inner_catalogue}

\begin{table}
	\centering
	\caption{Inner planets in systems that host at least one confirmed outer giant (OG sample).
          Listed are the planet name, radius $R_{\rm p}$ and mass $M_{\rm p}$ with their $1\sigma$ uncertainty. The sample contains
          $8$ planets around $7$ stars.}
	\label{tab:OG}
	\begin{tabular}{lcc} 
\hline
Planet Name & $R_{\rm p}\,[R_{\oplus}]$ & $M_{\rm p}\,[M_{\oplus}]$ \\
\hline
TOI-4010 c & $5.93\pm0.11$ & $20.3\pm2.1$ \\
TOI-1736 b & $3.18\pm0.11$ & $12.3\pm1.4$ \\
TOI-1694 b & $5.34\pm0.13$ & $31.3\pm2.6$ \\
Kepler-94 b & $3.51\pm0.15$ & $10.8\pm1.4$ \\
pi Men c & $2.11\pm0.05$ & $4.3\pm0.7$ \\
HD 191939 c & $3.19\pm0.07$ & $8.0\pm1.0$ \\
HD 191939 b & $3.40\pm0.09$ & $9.6\pm0.9$ \\
HAT-P-11 b & $4.45\pm0.10$ & $26.7\pm2.2$ \\
\hline
	\end{tabular}
\end{table}

\begin{table}
	\centering
	\caption{Same information as Table~\ref{tab:OG} but for inner planets in systems with
          no confirmed outer giant (NG sample).  This subsample comprises
          $34$ planets around $28$ stars.}
	\label{tab:NG}
	\begin{tabular}{lcc} 
    \hline
Planet Name & $R_{\rm p}\,[R_{\oplus}]$ & $M_{\rm p}\,[M_{\oplus}]$ \\
\hline
TOI-763 c & $2.63\pm0.12$ & $9.3\pm1.0$ \\
TOI-763 b & $2.28\pm0.11$ & $9.8\pm0.8$ \\
TOI-669 b & $2.60\pm0.17$ & $9.8\pm1.5$ \\
TOI-431 d & $3.29\pm0.09$ & $9.9\pm1.5$ \\
TOI-3568 b & $5.30\pm0.27$ & $26.4\pm1.0$ \\
TOI-3493 b & $3.22\pm0.08$ & $9.0\pm1.2$ \\
TOI-238 c & $2.18\pm0.18$ & $6.7\pm1.1$ \\
TOI-2141 b & $3.05\pm0.23$ & $24.0\pm4.0$ \\
TOI-1824 b & $2.74\pm0.08$ & $16.0\pm2.6$ \\
TOI-1803 b & $2.99\pm0.08$ & $10.3\pm1.1$ \\
TOI-1439 b & $4.24\pm0.20$ & $38.5\pm5.7$ \\
TOI-1272 b & $4.13\pm0.19$ & $27.0\pm2.6$ \\
TOI-125 d & $2.93\pm0.17$ & $13.6\pm1.2$ \\
TOI-125 c & $2.76\pm0.10$ & $6.6\pm1.0$ \\
TOI-125 b & $2.73\pm0.07$ & $9.5\pm0.9$ \\
TOI-1246 e & $3.78\pm0.16$ & $14.8\pm2.3$ \\
TOI-1246 c & $2.45\pm0.11$ & $9.1\pm1.2$ \\
TOI-1246 b & $3.01\pm0.06$ & $8.1\pm1.1$ \\
TOI-1064 b & $2.59\pm0.04$ & $13.5\pm1.8$ \\
TOI-1062 b & $2.27\pm0.09$ & $10.2\pm0.8$ \\
Kepler-19 b & $2.21\pm0.05$ & $20.3\pm2.8$ \\
Kepler-18 c & $5.22\pm0.35$ & $18.4\pm2.3$ \\
Kepler-11 d & $3.30\pm0.20$ & $6.8\pm0.8$ \\
Kepler-10 c & $2.35\pm0.02$ & $11.3\pm1.2$ \\
K2-285 c & $3.53\pm0.08$ & $15.7\pm2.2$ \\
K2-285 b & $2.59\pm0.06$ & $9.7\pm1.3$ \\
HIP 97166 b & $2.74\pm0.13$ & $20.0\pm1.5$ \\
HD 77946 b & $2.71\pm0.08$ & $8.4\pm1.3$ \\
HD 63935 b & $2.92\pm0.09$ & $10.5\pm1.6$ \\
HD 42813 b & $3.41\pm0.05$ & $9.1\pm0.8$ \\
HD 219666 b & $4.71\pm0.17$ & $16.6\pm1.3$ \\
HD 207897 b & $2.50\pm0.08$ & $14.4\pm1.6$ \\
GJ 143 b & $2.61\pm0.17$ & $22.7\pm2.0$ \\
EPIC 229004835 b & $2.33\pm0.09$ & $10.4\pm1.6$ \\
\hline
\end{tabular}
\end{table}

\bsp	
\label{lastpage}
\end{document}